# Spatial heterodyne scanning laser confocal holographic microscopy

## CHANGGENG LIU*

*Department of radiology and biomedical imaging, Yale school of medicine, New Haven, CT, 06510.(email: changgengliu34@gmail.com)*
*\*Present address: Department of Bioengineering, University of Illinois at Chicago.*

**Scanning laser confocal holographic microscopy using a spatial heterodyne detection method is presented. Spatial heterodyne detection technique employs a Mach-Zehnder interferometer with the reference beam frequency shifted by two acousto-optic modulators (AOM) relative to the object beam frequency. Different from the traditional temporal heterodyne detection technique in which hundreds temporal samples are taken at each scanning point to achieve the complex signal, the spatial heterodyne detection technique generates spatial interference fringes by use of a linear tempo-spatial relation provided by galvanometer scanning in a typical line-scanning confocal microscope or for the slow-scanning on one dimension in a point-scanning confocal microscope, thereby significantly reducing sampling rate and increasing the signal to noise ratio under the same illumination compared to the traditional temporal heterodyne counterpart. The proposed spatial heterodyne detection scheme applies to both line-scanning and point-scanning confocal microscopes. In this paper, we present the mathematical principles of the spatial heterodyne detection method and experimental schemes for both line-scanning and point-scanning confocal microscopes. Computer experiments are provided to demonstrate the validity of this idea. The presented spatial heterodyne scanning laser confocal holographic microscope (SH-SLCHM) can obtain both amplitude and quantitative phase information of the object field without a significant increase in complexity of optical and electronic design on the basis of a traditional scanning laser confocal microscope (SLCM). SH-SLCHM may find applications in optical metrology, in-vivo tissue imaging, and ophthalmic imaging.**

## 1. Introduction

Point-scanning confocal microscopy (PCM), invented in 1950s by Marvin Minsky [1, 2], has found wide applications in industrial inspection, biology, and medicine because it is capable of achieving high-resolution, high-contrast images, and true optical sectioning [3-5]. However, speed of PCM is limited by the point-scanning configuration. To increase the speed of PCM, parallel detection has been developed mainly in two modalities. The first is Nipkow spinning-disk confocal microscopy (SDCM) in which many points of the sample are simultaneously illuminated and detected by an area sensor [6]. Original SDCM using a double-sided Nipkow spinning disk configuration which suffers from low light illumination efficiency and difficulty in system alignment has been significantly improved by use of a single-sided Nipkow spinning disk configuration and a microlens [7, 8]. The second modality is line-scanning confocal microscopy (LCM), in which the sample is illuminated one line at a time, and the line field reflected or scattered from the sample is detected originally by film which has been replaced by digital cameras [9, 10]. LCM has a much simpler optical system compared to the PCM and SDCM, and can be engineered into a compact hand-held medical device [11, 12]. As a result, LCM has found wide application for ophthalmic imaging and tissue imaging [13-15].

Traditional confocal microscopy is an intensity-based imaging modality, which means the phase information of the sample is lost. To recover the phase information, interference methods have been applied to confocal microscopy by use of differential interference phase-contrast method, and electro-optical phase modulation [5]. Recently, digital holography (DH), an emerging quantitative phase-contrast imaging modality, has been widely studied and applied to optical metrology, biology, and medicine [16-18]. DH, as a wide-field imaging modality, can not reject the out-of-focus light to achieve true optical sectioning although digital refocusing is possible. Also it is a coherent imaging technique, in which speckle noise limits its use in tissue imaging. The concept, confocal holographic microscopy (CHM), was proposed by Herring [19, 20] to study fluids under microgravity. The original CHM working in transmission requires a very complicated optical system, which makes it hard to find applications in other fields. In 2012, a digital point-scanning confocal microscope (DPCM) was proposed by incorporating DH into PCM [21], and phase images of biological samples were reported in 2013 [22]. DPCM records the scattered or reflected light from one single point in the sample by a 2D digital hologram, which contains the information of the sample and the optical system. However, the speed is highly limited by this recording scheme. To speed up the imaging acquisition, we have recently reported digital line-scanning confocal holographic microscope (DLCHM), which has reduced the data flow by 3 orders of magnitude compared to DPCM, is able to achieve quantitative phase imaging at nanometer precision, and allows aberration compensation to improve the imaging resolution [23, 24]. The speed of DLCHM is still quite limited because each line field is recorded by a 2D digital hologram and ~1000 2D digital holograms are needed to reconstruct one en-face complex confocal image. Driven by the need of phase information from a near-field microscope, synthetic optical holography (SOH) was recently proposed [25], which has demonstrated the possibility of achieving one en-face complex confocal image from one single-shot spatial hologram by moving a reference beam. Currently, its image acquisition speed is limited at ~ 0.01 frame/s by the stage-scanning configuration, and a moving reference beam is not a practical choice for in-vivo imaging applications. Our recent development effort in digital adaptive optics ophthalmoscopy based on the principles of DH [24, 26, 27] to eliminate the hardware pieces and complicated close-loop feedback operation required by the hardware adaptive optics ophthalmoscopy[28, 29] has driven us to look for a feasible technical solution to obtain complex confocal images with beam-scanning configuration to achieve necessary frame rate (faster than 10frames/s) and without significantly increasing the complexity of the optical system and requirement of data acquisition on the basis of traditional point-scanning and line-scanning confocal ophthalmoscopy. In this paper, we present a spatial heterodyne detection method to achieve en-face complex confocal images for LCM and PCM, which can satisfy these requirements for in-vivo imaging applications. The proposed spatial heterodyne scanning laser confocal holographic microscopy (SH-SLCHM) is able to generate off-axis spatial hologram by use of a Mach-Zehnder interferometer with the frequency of the reference beam shifted by two acousto-optic modulators (AOM) relative to that of the object beam. Thanks to the linear tempo-spatial relation provided by the linear scanner such as Galvanometer scanning mirror (GSM) in a traditional LCM or for the slow scanning on one dimension in a traditional PCM, the phase ramp generated by the frequency shift in time leads to spatial interference fringes. SH-SLCHM can apply to both line-scanning and point-scanning confocal microscopes. For line-scanning SH-SLCHM (L-SH-SLCHM), the required frequency shift to achieve 40frames/s with an en-face images of 500×500 pixels is ~7500Hz. For point-scanning SH-SLCHM (P-SH-SLCHM), the corresponding frequency shift to achieve 10frames/s with en-face images of 500×500 pixels is ~1900Hz. In fact, use of two AOMs to generate a relative frequency shift between reference and object fields has been well researched in DH [30-32], where a frequency shift of several Hz is generated by two AOMs to obtain accurate phase-shifted holograms to achieve the complex field. Their work has demonstrated that it is possible to realize accurate frequency shift as low as several Hertz to milliHertz. High degree of accuracy of the frequency shift is required by phase-shifting interferometry. It is, however, not required by the proposed SH-SLCHM because the frequency shift is just introducing a proper spatial carrier frequency, from which a slight deviation will at most give rise to a linear phase ramp that can be numerically corrected in the data processing. Therefore, the frequency shift required by SH-SLCHM poses no technical issue. As we will present later, the fringe density of the generated off-axis digital holograms from SH-SLCHM can be easily controlled by setting different frequency shifts while keeping an in-line optical configuration. This is an advantage compared to off-axis wide-field DH and holographic projection of computer generated holograms, where the fringe density is often limited by the hardware [33-36]. To alleviate the requirement on the off-axis angle between the reference and object fields, different numerical methods have been proposed to remove the zero-order diffraction term in the reconstructed image plane in DH [33] or in the projection plane of computer generated holograms [34-36].

This paper is organized as follows. In section 2, the proposed optical system, mathematical principles and computer experiments for L-SH-SLCHM are presented. P-SH-SLCHMs with a linear fast scanner and a resonant fast scanner are then presented and discussed in section 3. In section 4, we will analyze the effect of nonlinearity of the tempo-spatial relation on the sampling rate, discuss the correction of possible phase aberration, and give a comparison of signal to noise (SNR) between the proposed spatial heterodyne method and the traditional temporal heterodyne method. Finally conclusions are drawn in section 4.

## 2. Line-scanning laser confocal holographic microscope

### A. Optical system

The optical system for the L-SH-SLCHM is depicted in Fig. 1. Figure 1(a) shows the top view of the optical layout. Assume that a He-Ne laser with central wavelength $\lambda$ 632.8nm is used as the light source. A cylindrical lens CL1 with a focal length of 150mm is employed to deliver a line focus oriented along y direction at the back focal plane of the microscope objective MO1 with a N.A. 0.65 and an effective focal length 4mm where the sample S is placed (Refer to the coordinates system in Fig. 1(a)). The line illumination through the cylindrical lens CL1 and the microscope objective MO1 is detailed in Fig. 1(b). The final image of the sample S is formed at the back focal plane of the lens L1 (focal length: 400mm), which results in a magnification of ~100 between the image plane and the sample plane. The line-scan CCD camera is put at the image plane, which has 500 pixels with a pixel pitch of 12μm(y direction) by 12μm(x direction) and a maximum line rate of 20KHz. The sample is scanned along x direction by use of a galvanometer scanning mirror GSM. To realize spatial heterodyne detection, a Mach-Zehnder interferometer is built by introducing a reference beam. The optical frequency of the reference field is shifted by $\Delta f$ using a pair of acousto-optic modulators AOM1 and AOM2. $\Delta f$ can be controlled by the frequencies of the driving signals. After the AOM2, the beam is collimated by the collimator C2, goes through the matching cylindrical lens CL2 with the same focal length as CL1, and is finally brought into focus at the back focal plane of a matching microscope objective MO2 of the same parameters as MO1. A mirror M is placed at the back focal plane of MO2. The reference focal line on the Mirror M is then magnified onto the line-scan camera by a factor of 100 using a matching imaging lens L2 with the same focal length as L1. The reference and object line fields arrive in parallel at the line-scan CCD camera sensor as shown in Fig. 1(a), resulting in 1D line-by-line in-line holograms along y direction. As we will describe in section 2B, the off-axis detection is realized along the scanning direction or x direction due to the optical frequency difference between the reference and object beams.

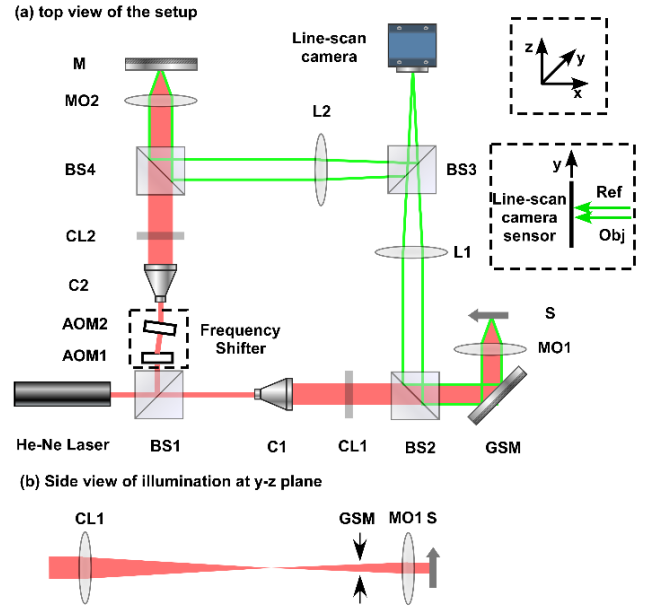

Fig. 1 Optical system of L-SH-SLCHM. (a) Top view of the optical layout. BS1-BS4: Beamsplitters. C1 and C2: Collimators. GSM: Galvanometer scanning mirror. MO1 and MO2: Microscope objectives. S: sample. L1 and L2: Imaging lens. AOM1 and AOM2: Acousto-optic modulators. M: Mirror. (b) Illumination on the sample at the y-z plane.

### B. Principle of operation

The line digital hologram at a given time t from this L-SH-SLCHM can be expressed as follows,

$$H(t, y) = |R(y)\exp(-j2\pi \Delta f\, t) + O(t, y)|^2$$
$$= |R(y)|^2 + |O(t, y)|^2$$
$$+ 2|R(y)||O(t, y)|\cos[2\pi \Delta f\, t + \varphi(t, y) + \varphi_A(t, y)] \quad (1)$$

where y is the spatial coordinate along the parallel detection, R(y) means the complex reference field at the camera plane, O(t, y) denotes the complex object field, $\varphi(t, y)$ is the phase information of the object field, and $\varphi_A(t, y)$ represents the phase aberration which contains the aberration of the optical system, and the phase jitter from mechanical vibrations. $\Delta f$ means the optical frequency shift of the reference field relative to the object field, which is generated by the two AOMs. Assume that the amplitude of the reference line field at the line-scan camera does not change with time t, which means R(y) is independent of the time t. Since the optical system employs an in-line configuration, the line hologram along y direction at a given time t is a 1D in-line hologram. Therefore we are not able to separate the zeroth term, image term and the twin image term in the 1D spatial Fourier Domain. However, we have introduced an optical frequency difference between the reference and the object beams. We can consider the resulting hologram as composed of 1D line-by-line holograms along the temporal axis which is linearly coupled with the spatial coordinate x along the scanning direction by the following relation:

$$x = v_x t \quad (2)$$

where $v_x$ is the speed at which the scanning line moves on the sample along x direction. Plugging Eq. (2) into Eq. (1), we can rewrite the resulting hologram at a given y position as

$$H(x, y) = |R(y)|^2 + |O(x, y)|^2 + 2|R(y)||O(x, y)|$$
$$\times \cos[2\pi f_{xc} x + \varphi(x, y) + \varphi_A(x, y)] \quad (3)$$

where

$$f_{xc} = \frac{\Delta f}{v_x} \quad (4)$$

Eq. (3) tells us that the final hologram is composed of 1D off-axis holograms along x or the scanning direction due to the tempo-spatial relation of the time and the spatial coordinate x as described by Eq.(2). From Eq. (4), the spatial carrier frequency $f_{xc}$ or the fringe density of the hologram can be adjusted by changing $\Delta f$. The sampling spacing dx along the scanning direction is restricted by the N.A. of MO1 and the wavelength $\lambda$. Considering that the lateral resolution of L-SH-SLCHM is very close to that of the wide-field coherent imaging system, we can use the frequency analysis for wide-field coherent imaging system to estimate the sampling spacing requirement. According to ref. [37], the cutoff spatial frequency $f_{cutoff}$ of the optical system can be given by $f_{cutoff} = \frac{N.A.}{\lambda}$. That means the widths of the spatial frequency of image term, twin-image term and zeroth-order term are $\frac{2N.A.}{\lambda}$, $\frac{2N.A.}{\lambda}$, and $\frac{4N.A.}{\lambda}$ respectively. To avoid aliasing, the total extent of the spatial frequency from the digital hologram that is given by $\frac{1}{dx}$ has to contain the total frequency extent of these three terms which is $\frac{8N.A.}{\lambda}$. Therefore the sampling spacing dx has to satisfy that $dx \leq \frac{\lambda}{8N.A.}$. In the proposed setup, dx should be finer than 0.122μm. This is a prerequisite to achieve the phase and amplitude without cross-talk noise from the holographic process no matter how densely the interference fringes can be set. dx can be larger if the inner spatial frequency bandwidth of the object field is narrower than double cutoff spatial frequency $f_{cutoff}$ of the optical system. To demonstrate the principle of the operation, a computer experiment will be presented in section 2C.

### C. Computer simulations

The experimental parameters are set as follows. We set the line period dt of the line-scan camera to be 50 μs. The magnification of the optical system is 100. The sampling spacing dy along y direction is computed by the ratio of the pixel pitch (12μm) over the magnification 100 as 0.12μm. To be consistent with dy, dx is also set to be 0.12μm, which is slightly finer than the sampling requirement 0.122μm. To separate the three terms in the spatial Fourier domain, the spatial carrier frequency $f_{xc}$ or the spatial fringe density along x direction is set to be 0.37 cycles/pixel or 3.12cycles/μm. The speed $v_x$ can be calculated by dx/dt as 2400μm/s. From Eq. (4), $\Delta f$ can be calculated as 7480Hz. A complex image of 500×500 pixels that corresponds to a field of view (FOV) ~60×60μm² is used in the computer experiment, which results in a frame rate of 40 frames/s. For simplicity, the phase aberration term in Eq. (3) is set to be zero and the amplitude of the reference line field R(y)=1. To provide the correspondence between the virtual computer experiment and the physical reality, the highest spatial frequency of the complex confocal image in use should not exceed the cut-off spatial frequency $f_{cutoff}$ which is estimated by $\frac{N.A.}{\lambda}$ as 1.03cylces/μm or ~61 pixels in the spatial frequency domain. This is the key that makes this computer experiment meaningful. The amplitude and phase map (in radian throughout the remainder of this manuscript) of the complex image is shown in Figs. 2(a) and (b) respectively. The resulting hologram is shown in Fig. 2(c). The detailed view of a portion of Fig. 2(c) is shown in Fig. 2(d). We first present a widely used 2D method to reconstruct the complex image from the hologram. The 2D angular spectrum (AS) of Fig. 2(c) as shown in Fig. 2(e) contains the image term (highlighted by the green circle), the zeroth term at the center, and the twin-image term at the right. The image term is then filtered out, shifted to the center and inversely Fourier transformed. The resulting reconstructed amplitude and phase map are shown in Figs. 2(f) and 2(g) respectively. We then present a 1D method which is indeed equivalent to the 2D method and will become a more general method to process scanning confocal holographic data. Instead of treating the hologram as 2D hologram as a whole, we can decompose it into 500 1D off-axis line-by-line holograms along x direction. This can be seen from Eq. (3). For a given y position, the equation represents a 1D hologram along x direction. We can process the holographic data in a line-by-line manner and combine the reconstructed line fields into the final en-face image. The 1D AS of Fig. 2(c), which is obtained by taking 1D FT of it along x direction, as shown in Fig. 2(h) also contains the image term, zeroth term and twin-image term. The image term is bounded by the green rectangle in Fig. 2(h). The profile along the white dashed lines in Fig. 2(h) is shown in Fig. 2(i). We can reconstruct the complex field in a line-by-line manner. The resulting reconstructed amplitude and phase map obtained by the 1D method as shown in Figs. 2(j) and 2(k) are the same as the reconstructions by the 2D method, since both methods are equivalent if the sampling requirement is met and the fringe density is sufficient along x direction. To further validate effectiveness and equivalence of the 2D and 1D methods, we then present a more general simulation case where amplitude and phase map appears completely different. The simulated experimental parameters remain unchanged. The amplitude and phase map of this simulated complex image are shown in Figs. 2(l) and 2(m). The corresponding hologram and its detailed view are shown in Figs. 2(n) and 2(o). The reconstructed amplitude and phase map by the 2D method are shown in Figs. 2(p) and 2(q). The amplitude and phase map reconstructed by the 1D method are shown in Figs. 2(r) and 2(s). As we can see, the reconstructions from the 1D method and 2D method are the same, which indicates that the 2D and 1D methods are effective and equivalent. The results from these two simulated experiments also show that the complex field can always be recovered if the sampling spacing is fine enough and the interference fringes are dense enough to realize complete separation of the three terms in the AS domain.

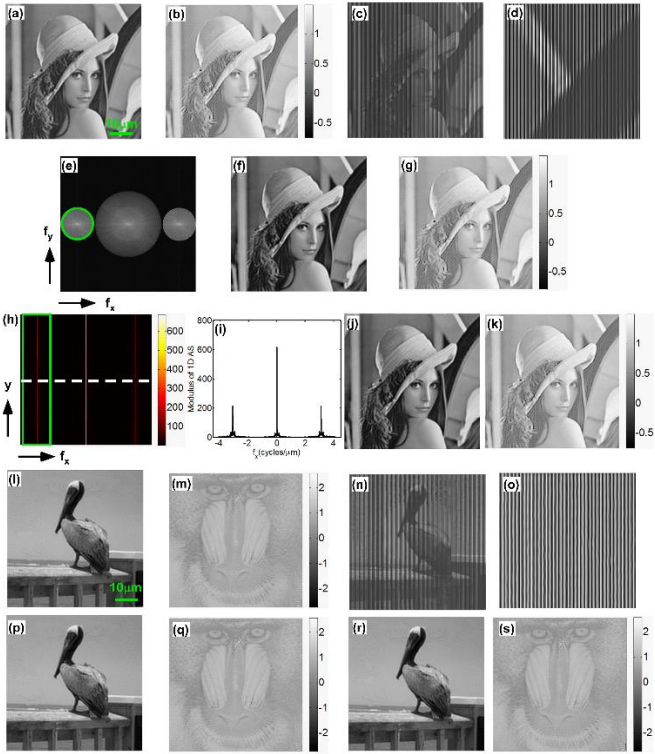

Fig. 2 Computer simulation for L-SH-SLCHM. (a) Amplitude of simulated image, (b) Phase map (in radian throughout the manuscript) of the simulated image. (c) Spatial heterodyne hologram. (d) Detailed view of (c). (e) 2D AS in logarithmic amplitude scale. (f) Reconstructed amplitude from the 2D method. (g) Reconstructed phase from the 2D method. (h) Modulus of 1D AS in linear amplitude scale. (i) Profile along the white dashed line in (h). (j) Reconstructed amplitude from the 1D method. (k) Reconstructed phase from the 1D method. (l) and (m) are the amplitude and phase map of a second simulated complex image for L-SH-SLCHM. (n) Spatial heterodyne hologram of the complex image represented by (l) and (m). (o) Detailed view of the (n). (p) and (q) are the reconstructed amplitude and phase map by the 2D method. (r) and (s) are the reconstructed amplitude and phase map by the 1D method.

## 3. Point-scanning laser confocal holographic microscope

### A. Optical system

The optical system for the P-SH-SLCHM is shown in Fig. 3. MO1 and MO2 are the same as in Fig. 1(a). The regular lens L1 and L2 constitute a 1:1 imaging system between the galvanometer scanning mirror GSM and the Polygonal or Resonant scanning mirror, and lens L3 and L4 forms a 1:1 imaging system between the pupil plane of MO1 and the galvanometer scanning mirror GSM. The illumination is focused at the back focal plane of MO1 where the sample S is placed. The imaging lens L5 is set to be 400mm in focal length so that magnification between imaging plane and the sample plane is 100. A pinhole with a diffraction-limited resolution width is placed at the final image plane. On the reference end, a quite symmetrical configuration is suggested here. The matching imaging lens L6 of the same focal length of L5 will bring the light into focus at the pinhole plane. The interference signal will be detected by a detector which is placed immediately after the pinhole. We assume that the slow scanning (x direction) is achieved by the GSM, which is widely adopted for the slow scanning on one dimension in the traditional PCM. Since GSM can provide high degree of linearity of scanning, it is reasonable to assume a linear tempo-spatial relation between spatial coordinate x and tx. tx is the temporal lapse along x direction at a given y position. For the fast scanning axis (y direction), we will discuss two cases. The first one is linear scanning which can be realized by a spinning polygonal mirror but not widely used. The second one which is a widely used scheme is resonant scanning provided by a resonant scanner. For both cases, we set the frame rate to be 10 frames/s with a frame of N×N pixels where N is set to be 500.

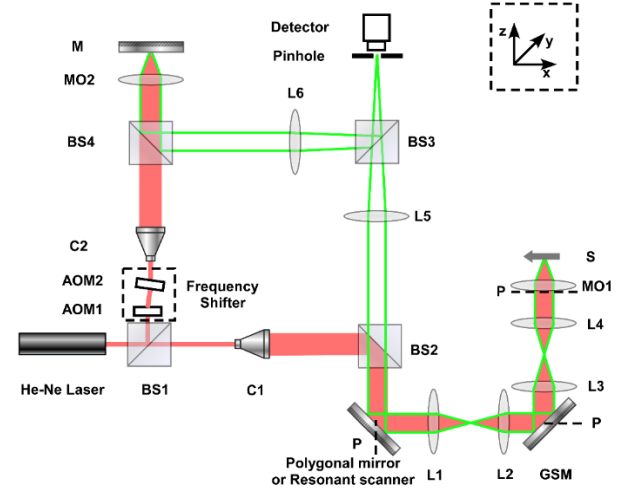

Fig. 3 Optical system for P-SH-SLCHM. BS1-BS4: Beamsplitters. C1-C2: Beam collimators. L1-L6: Regular lens. GSM: Galvanometer scanning mirror. MO1-MO2: Microscope objectives. P: Pupil plane or its conjugate plane. S: sample. AOM1-AOM2: Acousto-optic modulators. M: Mirror.

### B. Principle of operation for linear fast scanning

In this section, we present the mathematical principle of P-SH-SLCHM with a linear fast scanner such as spinning polygonal mirror. The digital hologram can be expressed as

$$H(t) = |R|^2 + |O(t)|^2 + 2|R||O(t)| \times \cos[2\pi \Delta f\, t + \varphi(t) + \varphi_A(t)] \quad (5)$$

where t is the time lapse relative to the starting point of a frame. $\varphi(t)$ is the phase information of the object, $\varphi_A(t)$ represents the phase aberration term as described in section 2B. $\Delta f$ represents the optical frequency shift of the reference field R relative to the object field O(t), which is realized by the two AOMs. Since the amplitude of the reference point field does not change with time, R is independent of the time t. The spatial sampling grids are illustrated in Fig. 4(a). We assume that scanning starts from the origin (0, 0) at the sample plane, goes upward along the positive direction of y axis, and then goes back to (dx, 0) without retracing on the sample. This way the sample is raster scanned without backward scanning along the negative direction of y axis. Also the time delay between the end of a fast scanning line along y direction and the starting point of the next new fast scanning line is ignored for simplicity. The scanning paths on x and y directions are approximately orthogonal because the scanning speed $v_y$ along y direction is 499 times faster than the scanning speed $v_x$ along x direction if the sampling spacing dx and dy along the two directions are set to be the same, which is a practical choice to avoid image distortion. The time t can be decomposed into

$$t = t_x + t_y \qquad (6)$$

where $t_x$ represents the time needed to reach the starting point of a vertical line, and $t_y$ denotes the time lapse along y direction relative to the starting point of this specific vertical line. The spatial coordinates (x, y) that correspond to the time t are illustrated by the point A in Fig .4(a). $t_x$ and $t_y$ can be given by

$$t_x = \frac{x}{v_x}, \qquad t_y = \frac{y}{v_y} \qquad (7)$$

To be clear, the relationship between y and $t_y$ is illustrated in Fig. 4(b), where $dt_y$ means the temporal sampling interval along y direction, and dy denotes the spatial sampling spacing along y direction. Plugging Eqs. (6) and (7) into the Eq.(5), we can obtain a spatial digital hologram, as follows,

$$H(x, y) = |R|^2 + |O(x, y)|^2$$
$$+ 2|R||O(x, y)|\cos[2\pi(f_{xc} x + f_{yc} y) + \varphi(x, y) + \varphi_A(x, y)] \qquad (8)$$

where $f_{xc}$ and $f_{yc}$ are the spatial carrier frequencies along x and y directions respectively, which are given by

$$f_{xc} = \frac{\Delta f}{v_x}, \qquad f_{yc} = \frac{\Delta f}{v_y} \qquad (9)$$

Both $f_{xc}$ and $f_{yc}$ are proportional to the frequency shift $\Delta f$ and inversely proportional to their respective scanning speeds. From Eq. (9), we can obtain $f_{xc} = \frac{v_y}{v_x} f_{yc}$, which means $f_{xc}$ is ~N times higher than $f_{yc}$. In practice, sampling spacing dx and dy are set to be the same to avoid image distortion, and need satisfy the sampling requirement by off-axis direction. Compared to L-SH-SLCHM, the sampling requirement for P-SH-SLCHM is higher. The coherent point spread function (PSF) of P-SH-SLCHM is the product of the coherent illumination PSF and imaging PSF. That means the transfer function of P-SH-SLCHM is the convolution of that from coherent wide field imaging system. Therefore the cutoff spatial frequency of P-SH-SLCHM is doubled compared to L-SH-SLCHM. Similar to the analysis provided in section 2B, the sampling requirement for P-SH-SLCHM becomes $dx \leq \frac{\lambda}{16N.A.}$. This is a quite conservative estimate because the frequency content for P-SH-SLCHM will fall off at the high frequency end because the profile of the transfer function along one direction is like a triangle function instead of a rectangle function. For simplicity, in our simulations, we will adopt this conservative estimate to set the sampling spacing. The corresponding temporal sampling interval $dt_x$ and $dt_y$ along x and y directions can be computed by the information on frame rate and size. $dt_x$ is related to $dt_y$ by $dt_x = Ndt_y$. We can then calculate the speed $v_x$ and $v_y$ by Eq. (7). $\Delta f$ is chosen so that $f_{xc}=0.37$ cycles/pixel to ensure separation of three terms in the spatial Fourier domain. Since $f_{xc}=\sim500$ $f_{yc}$, we can only obtain either vertical or horizontal interference fringes as will be demonstrated later in section 3C. To reconstruct the complex signal, we will adopt the 1D method to process the resulting hologram which means we treat it as composed of 500 1D off-axis line-by-line hologram along x or y direction as described in section 2C. This way, the spatial sampling requirement along the holographic direction is decoupled with the sampling spacing on the non-holographic direction. The resulting complex signal contains a phase ramp along x or y direction, given by $\exp(2\pi j f_{xc} x)$ or $\exp(2\pi j f_{yc} y)$, where j denotes $\sqrt{-1}$.

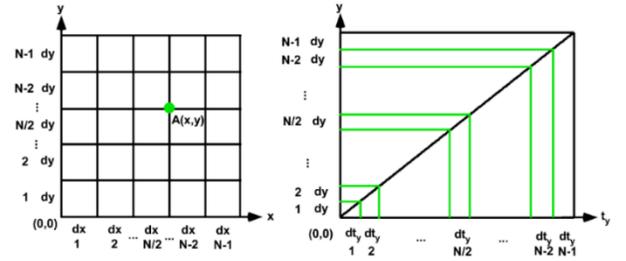

Fig. 4 Spatial sampling grids and scanning pathway with $t_y$ along y direction for the linear fast scanning configuration. (a) Spatial sampling grids. (b) The dependence of y on $t_y$.

### C. Computer simulations for linear fast scanning

We will present two computer experiments. In the first experiment, we will set experimental parameters so that the fringe density along x direction satisfies the separation condition in spatial Fourier domain. We set dx=dy=0.06μm, and frame rate to be 10 frames/s with an en-face image of 500×500 pixels. Therefore, the $dt_y$ is 0.4μs and $dt_x=500dt_y=200$μs. The speed $v_x=dx/dt_x=300$μm/s and $v_y=500v_x=150$mm/s. The fringe density along x direction is then set to 0.37 cycles/pixel or 6.24 cycles/μm. $f_{yc}$ is then computed as 0.013 cycles/μm. $\Delta f$ for this case is computed by $f_{xc}v_x$ as 1870Hz. The complex image is same as shown in Figs. 2(a) and (b) where both dx and dy are 0.06μm. The cut-off spatial frequency for this complex image is doubled because the sampling spacing is halved compared to L-SH-SLCHM, which is just the required one for P-SH-SLCHM. The resulting hologram from this experiment is shown in Fig. 5(a). The reconstructed amplitude and phase map are shown in Figs. 5(b) and (c) by 1D method as demonstrated for L-SH-SLCHM. The phase map in Fig. 5(c) shows a phase ramp given by $\exp(2\pi j f_{yc} y)$. After correction of this ramp, the resulting corrected phase map is shown in Fig. 5(d). In the second experiment, we set $f_{yc}$ to be 6.24 cycles/μm. $\Delta f$ can be then computed by $f_{yc}v_y$ as high as 0.935MHz and the corresponding $f_{xc}$ becomes 3116.7 cycles/μm. With such a high spatial carrier frequency along x direction, the fringes along this direction are essentially washed out. Therefore, the resulting hologram contains only horizontal fringes as shown in Fig. 5(e). Since $f_{yc}$ has separated the three term in the spatial Fourier domain, the complex amplitude of the image is able to be achieved. The reconstructed amplitude is shown in Fig. 5(f) and the phase map before the correction of the phase ramp $\exp(2\pi j f_{xc} x)$ is shown in Fig. 5(g). The corrected phase map is shown in Fig. 5(h). In fact, due to the large spatial carrier frequency along x direction, each pixel has averaged out hundreds cycles, resulting in no observable phase ramp.

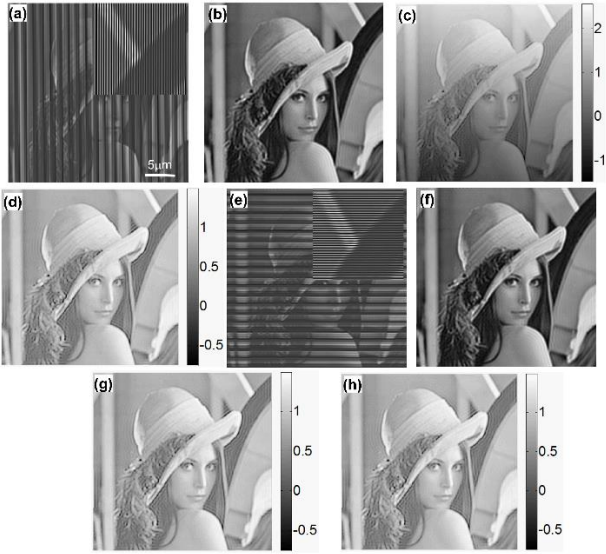

Fig. 5 Demonstration for P-SH-SLCHM with a linear fast-scanning configuration. (a) Hologram with Δf=1870Hz (Inset 2.5×). (b) Reconstructed amplitude from (a). (c) Reconstructed phase map from (a) with a phase ramp along y direction. (d) Phase map after the phase ramp correction. (e) Hologram with Δf=0.935MHz (Inset 2.5×). (f) Reconstructed amplitude from (e). (g) Reconstructed phase map before phase ramp correction from (f). (h) Corrected phase map from (g).

### D. Principle of operation for resonant fast scanning

When we use a resonant scanner in the fast-scanning axis and a linear scanner in the slow-scanning axis as a typical traditional PCM adopts, the sampling spacing along y direction is not uniform. The resulting sampling grids are illustrated by Fig. 6(a), where the sampling spacing along y direction in the center is much wider than that on the two sides. In this case, only forward scanning lines along the positive direction of y axis are shown in Fig. 6(a). The backward scanning lines along the negative direction of y axis are ignored because the signal recovery process is similar as the forward scanning lines. The spatial coordinate y is related to $t_y$ by

$$y = \frac{L_y}{2}\{1 - \cos[\frac{\pi t_y}{(N-1)dt_y}]\},$$
$$t_y = \{0, \ dt_y, \ ..., \ (N-2)dt_y, \ (N-1)dt_y\} \quad (10)$$

where $L_y$ is full FOV along y direction, $dt_y$ is the temporal sampling interval along y direction. This equation is illustrated in Fig. 6(b), where we can see that the sampling spacing along y direction varies at the different temporal instants $t_y$. From Eq. (10), $t_y$ can be written as

$$t_y = \frac{(N-1)dt_y}{\pi}\cos^{-1}(1-\frac{2y}{L_y}) \quad (11)$$

where $\cos^{-1}(\cdot)$ represents inverse cosine function. The time t that corresponds to a given point B(x, y) as shown in Fig. 6(a) is still governed by Eq. (6). Note that $dt_x$ is equal to $2Ndt_y$. $t_x$ is given by Eq. (7). The resulting digital hologram can be expressed by

$$H(x,y) = |R|^2 + |O(x,y)|^2 + 2|R||O(x,y)|\cos\{2\pi[f_{xc} x + \Delta f\frac{(N-1)dt_y}{\pi}\cos^{-1}(1-\frac{2y}{L_y})] + \varphi(x,y) + \varphi_A(x,y)\} \quad (12)$$

where $f_{xc}$ is given by Eq. (9). We can process this hologram as composed of N 1D line-by-line holograms along x direction. The reconstructed complex signal will contain a phase term by

$\exp\{2\pi j[\Delta f\frac{(N-1)dt_y}{\pi}\cos^{-1}(1-\frac{2y}{L_y})]\}$, which can be numerically corrected.

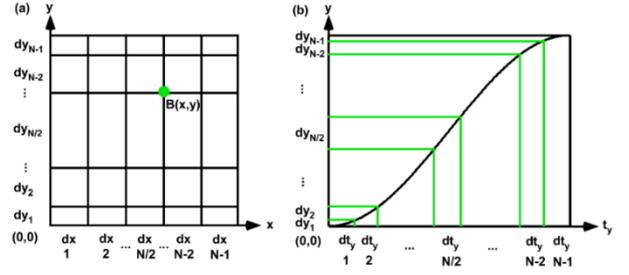

Fig. 6 Spatial sampling grids and resonant scanning pathway with $t_y$ along y direction for the resonant fast scanning configuration. (a) Spatial sampling grids. (b) The dependence of y on $t_y$.

### E. Computer simulations for resonant fast scanning

In this computer experiment, we set dx=0.06μm and average sampling spacing dy along y direction to be 0.06μm. We set the frame rate to be 10 frames/rate with 500×500 en-face image (500 backward scanning lines are ignored), which corresponds to a FOV of ~30×30μm². $dt_y$ is computed as 0.2μs, and $dt_x$ can then be calculated by $dt_x$=1000$dt_y$ as 200μs. The speed $v_x$ can be computed by dx/$dt_x$ as 300μm/s and the average speed along y direction $\overline{v_y}$ =1000$v_x$=300mm/s. The fringe density along x direction is set to be 6.24cycles/μm. Δf is computed by the product of $f_{xc}$ and $v_x$ as 1870Hz. We can compute the actual sampling positions y(n)={y(0), y(1), ... y(498), y(499)} with n={0, 1, ... 498, 499} at different time $t_y$ by Eq. (10). The original complex image before distortion due to the non-uniform sampling along y direction are shown in Figs. 2(a) and (b). Note that both dx and dy are set to be 0.06μm. To simulate the distortion due to the resonant scanning, we resample the original image by use of linear interpolation along y direction. Assume that $y_u$(n) with n={0,1,...498, 499} denotes the uniform sampling positions which are given by {0, dy ,..., 498dy, 499dy}, and O(y) denotes the complex image at a given x position. The resampling procedure in a row-by-row manner is described as follows. O[y(0)]= O[$y_u$(0)], and O[y(499)]= O[$y_u$(499)]. For n ranging from 2 to 498, calculate y(n)/dy=m+δ where m is the quotient and δ is the remainder, and set O[y(n)] to be (1-δ)O[$y_u$(m)]+ δO[$y_u$(m+1)]. The amplitude and phase map of the resulting distorted complex image are shown in Figs. 7(a) and (b) respectively. Since the distortion procedure is linear, it will not affect the spatial frequency bandwidth of the 1D hologram along x direction and the corresponding complex signal recovery. Figure 7(c) shows the corresponding digital hologram as described by Eq. (12) which consists of 500 1D off-axis holograms along x direction. We then process these holograms in a line-by-line manner. The resulting reconstructed amplitude and phase map are shown in Figs. 7(d) and 6(e) respectively, which indicates that the non-uniformity of the sampling spacing along y direction does not affect the recovery of the amplitude and phase of the complex signal, because sampling spacing dx is fine enough and separation condition is met by setting a proper Δf. To correct the distorted reconstructed image, we apply a similar linear interpolation algorithm as follows. O[$y_u$(0)]=O[y(0)], and O[$y_u$(499)]= O[y(499)]. For m ranging from 2 to 498, find closest y(n) for $y_u$(m), and calculate [$y_u$(m)-y(n)]/dy= δ. If $\delta \geq 0$, O[$y_u$(m)] is set to be (1-δ)O[y(n)]+ δO[y(n+1)]. If $\delta < 0$, O[$y_u$(m)] is given by (-δ)O[y(n-1)]+ (1+δ)O[y(n)]. The amplitude and phase map of the corrected complex reconstruction are shown in Figs. 7(f) and 7(g) respectively.

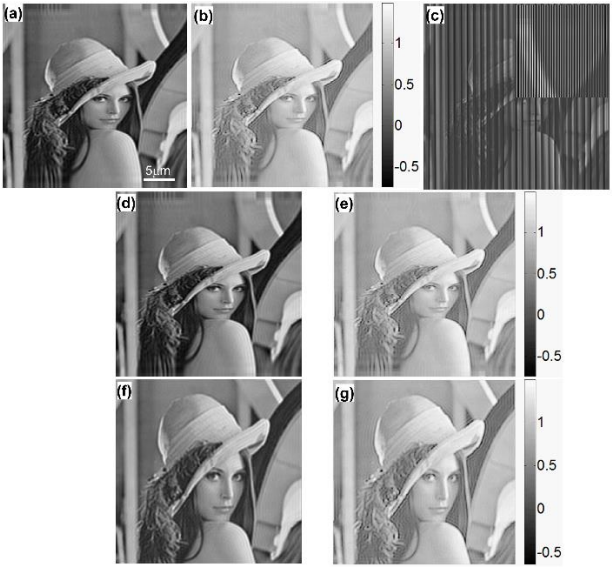

Fig. 7 Demonstration of P-SH-SLCHM with a resonant scanner. (a), and (b) Amplitude and phase map of the complex image. (c) Hologram (Inset 2.5×). (d), and (e) Reconstructed amplitude and phase map from (c). (f), and (g) Amplitude and phase map after distortion correction.

## 4. Discussions

### A. Effect of nonlinearity of the tempo-spatial relation on sampling rate

SH-SLCHM is based on the assumption that the GSM can provide a linear tempo-spatial relation. We will provide a quantitative analysis to assess the effectiveness of this assumption and the temporal sampling rate requirement if this temporal-spatial relation is not held well. A simplified diagram for the GSM scanning configuration is shown in Fig. 8, where θ is the scanning angle and f is the focal length of the microscope objective MO1 as described in Fig. 1(a). Assume that the beam is scanning from the left to the right with the scanning angle θ ranging from $-\theta_{max}$ to $+\theta_{max}$, where $\theta_{max}$ is the largest positive scanning angle. Assume that the angular speed $v_a$ (in radian/s) of the scanning beam is a constant. The spatial coordinate x at the point C as shown in Fig. 8 is related to the corresponding scanning angle θ by $x(\theta) = f\tan(\theta) + f\tan(\theta_{max})$. Under first-order approximation, the sampling spacing at θ can be expressed as $dx(\theta) = \frac{fd\theta}{\cos^2(\theta)} = \frac{dx}{\cos^2(\theta)}$ where dx is the sampling spacing at θ=0 degree and is the smallest sampling spacing. The scanning speed along x direction at θ can be given by $v_x(\theta) = \frac{v}{\cos^2(\theta)}$ where v is the speed at θ=0 degree and is the slowest speed of the scanning point moving at the sample. To give a quantitative sense of the nonlinearity of this scanning configuration, we can define the degree of nonlinearity by

$$D_{Non} = \frac{dx_{max} - dx}{dx} \times 100\% = [\frac{1}{\cos^2(\theta_{max})} - 1] \times 100\% \quad (13)$$

where $dx_{max}$ means the largest sampling spacing which occurs when $\theta = \pm\theta_{max}$. We can then evaluate the nonlinearity of the optical system we presented in this paper, where f=4mm and the FOV along x direction is 60μm for the L-SH-SLCHM. At this FOV, the largest scanning angle $\theta_{max}$ is computed as 0.43 degree. The corresponding degree of nonlinearity $D_{Non}$ is 0.006%. If we increase the FOV to 600μm along x direction, which corresponds to the $\theta_{max}$ of 4.3 degree. At this FOV, $D_{Non}$ is computed as ~0.6%, which means at this FOV, the linearity of the tempo-spatial relationship still holds very well. Therefore, under a normal FOV of a scanning confocal microscope, the tempo-spatial relationship can be considered as linear. From Eq. (13), $D_{Non}$ increases as the largest scanning angle $\theta_{max}$ increases. If the $\theta_{max}$ is large enough so that the linearity does not hold well anymore, there arises a question on determination of the sampling rate. To be specific, if the $\theta_{max}$ is set to be 45 degree which corresponds to a FOV of 8mm, $D_{Non}$ becomes 100%. At this FOV, the spatial sampling spacing at $\theta = \pm\theta_{max}$ is double the one at $\theta = 0$ degree. The spatial sampling spacing becomes very non-uniform. However, the temporal sampling interval $dt_x$ is still uniform. If the spatial frequency bandwidth of the complex signal O(x) for a given y position is $W_S$. The corresponding temporal frequency bandwidth $W_T$ of the complex signal $O_T(t) = O[x(t)]$ can be conservatively estimated by $W_T = W_S v_{x,max}$, where $v_{x,max} = 2v$ is the maximum speed along x direction. To meet the sampling requirement for off-axis detection, $dt_x$ has to satisfy $dt_x \leq \frac{1}{4W_T}$. Take L-SH-SLCHM as example. The spatial frequency bandwidth $W_S$ can be estimated by $W_S = \frac{2N.A.}{\lambda}$ and the corresponding temporal frequency bandwidth $W_T$ can then be estimated as $W_T = \frac{4vN.A.}{\lambda}$. Then the temporal sampling interval $dt_x$ has to satisfy that $dt_x \leq \frac{\lambda}{16vN.A.}$. In fact, this temporal sampling interval is obtained from a pessimistic point of view because it is equivalent to that we have to set the largest spatial sampling spacing $dx_{max}$ to satisfy that $dx_{max} \leq \frac{\lambda}{8N.A.}$. It needs further theoretical investigation to find the optimal temporal sampling interval, which may be set longer than our conservative estimate.

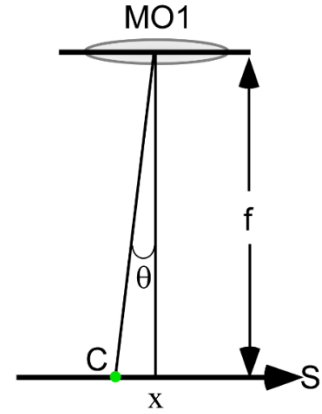

Fig. 8 The simplified diagram of the GSM scanning configuration. f represents the focal length of the MO1, and S means the sample plane as described in Fig. 1(a).

### B. Phase aberration correction

After we achieve the complex field from the spatial heterodyne hologram, the phase aberration term $\varphi_A(x, y)$ will have to be removed to clearly reveal the phase information of the sample. $\varphi_A(x, y)$ can be decomposed into two parts by $\varphi_A(x, y) = \varphi_{A,D}(x, y) + \varphi_{A,R}(x, y)$, where $\varphi_{A,D}(x, y)$ means deterministic phase aberration from the optical system and $\varphi_{A,R}(x, y)$ is the random phase variations from the mechanical vibration. The removal of $\varphi_{A,D}(x,y)$ has been widely studied in the field of DH. It can be numerically removed by Polynomial fitting procedures [18]. For line-scanning confocal imaging system, $\varphi_{A,R}(x,y)$ depends only on the x, and can be removed by a phase jitter correction method, which has been detailed in ref. [23]. For point-scanning confocal holographic microscope, we expect that the $\varphi_{A,R}(x,y)$ will also mainly depend on the x or slow-scanning direction, which has been experimentally validated from a stage-scanning SOH microscope where a moving reference beam is used to introduce the spatial fringes [25].

### C. Spatial heterodyne versus temporal heterodyne

Spatial heterodyne detection adopts a similar method used by temporal heterodyne detection [38, 39]. Both of them employ a local oscillator with a center frequency shifted relative to that of the signal to detect. Different from temporal heterodyne detection, spatial heterodyne detection imposes a tempo-spatial relationship, by which we convert temporal fringes into spatial fringes. This tempo-spatial relationship in turn couples the N.A. of the microscope objective with the sampling spacing. The sampling rate of temporal heterodyne detection is determined by the temporal frequency bandwidth of the temporal signal, which has nothing to do with the N.A. of the microscope objective in the object arm. The signal-to-noise ratio SNR for both methods under the condition that a much stronger reference field is employed compared to the object field can be expressed by [39]

$$SNR = 10\log(\frac{\eta |O|^2 \Delta A \Delta t}{hw}) \tag{14}$$

where $\eta$ means quantum efficiency of the detector, $|O|^2$ denotes the intensity of the object field at the detector, $\Delta A$ represents the detector area, $\Delta t$ is the exposure time, h is the Planck constant, and w is the frequency of the light source. For simplicity, we assume that the object is a uniform reflector and the illumination is the same for both temporal heterodyne and spatial heterodyne detections, then $|O|^2$ will be the same. To maintain the same pixel rate, the exposure time $\Delta t_{TH}$ for temporal heterodyne detection with M temporal samples will be ~M times shorter than the exposure time $\Delta t_{SH}$ for spatial heterodyne detection. M is usually more than 100, which means the SNR for temporal heterodyne detection will be ~20dB lower than spatial heterodyne detection.

## 5. Conclusions

In summary, we have presented a spatial heterodyne detection technique for both LCM and PCM which can achieve en-face complex confocal image of the sample by adding a reference field with a shifted optical frequency. The spatial heterodyne digital holograms are generated by the temporal phase ramp due to the optical frequency shift and the tempo-spatial relation produced by linear beam scanning configuration in LCM or along the slow-scanning axis in PCM. A general 1D data processing is described, by which the sampling requirement is imposed only on the holographic direction. This is why spatial heterodyne detection can be applied to PCM with a fast resonant scanner which gives rise to very non-uniform spatial sampling along the fast-scanning direction. The proposed setups for SH-SLCHM, which are in-line symmetric configurations, allow low-coherence light sources by use of low-coherence DH with careful adjustment of the path lengths of the two arms [40-43]. Capable of achieving quantitative phase information, SH-SLCHM may not only find applications for optical precision measurement and in-vivo tissue imaging, but also open the possibility of developing digital adaptive optics confocal ophthalmoscopes by use of measured aberration [24, 26, 27] or by numerical methods [44-46].

**Acknowledgment**. The author is grateful to his PhD advisor Prof. Myung K. Kim at University of South Florida for leading him to the research on digital holographic adaptive optics ophthalmoscopy which is the driving force to search for this idea. The author is also grateful to his postdoc supervisor Prof. Michael A. Choma at Yale School of Medicine for sharing his experience and knowledge in optical coherence tomography.